\input harvmac
\input epsf
\noblackbox
%-------------------------
% This paper uses harvmac
%-------------------------
\overfullrule=0pt
\def\Title#1#2{\rightline{#1}\ifx\answ\bigans\nopagenumbers\pageno0\vskip1in
\else\pageno1\vskip.8in\fi \centerline{\titlefont #2}\vskip .5in}
 
scaled\magstep3 
 
scaled\magstep3 
 
scaled\magstep3 
 
\font\cmss=cmss10 \font\cmsss=cmss10 at 7pt
%
%%%%%%%%%%%%%%%%%%%%%%%%%%  FIGURES   %%%%%%%%%%%%%%%%%%%%%%%%%%%%%%%

\newcount\figno
\figno=0
\def\fig#1#2#3{
\par\begingroup\parindent=0pt\leftskip=1cm\rightskip=1cm\parindent=0pt
\baselineskip=11pt \global\advance\figno by 1 \midinsert
\epsfxsize=#3 \centerline{\epsfbox{#2}} \vskip 12pt {\bf Fig.\
\the\figno: } #1\par
\endinsert\endgroup\par
}
\def\figlabel#1{\xdef#1{\the\figno}}
\def\encadremath#1{\vbox{\hrule\hbox{\vrule\kern8pt\vbox{\kern8pt
\hbox{$\displaystyle #1$}\kern8pt} \kern8pt\vrule}\hrule}}
%%%%%%%%%%%%%%%%%%%%%  Math-style letters   %%%%%%%%%%%%%%%%%%%%%%%%
\font\cmss=cmss10 \font\cmsss=cmss10 at 7pt

\def\IB{\relax\hbox{$\inbar\kern-.3em{\rm B}$}}
\def\IC{\relax\hbox{$\inbar\kern-.3em{\rm C}$}}
\def\IQ{\relax\hbox{$\inbar\kern-.3em{\rm Q}$}}
\def\ID{\relax\hbox{$\inbar\kern-.3em{\rm D}$}}
\def\IE{\relax\hbox{$\inbar\kern-.3em{\rm E}$}}
\def\IF{\relax\hbox{$\inbar\kern-.3em{\rm F}$}}
\def\IG{\relax\hbox{$\inbar\kern-.3em{\rm G}$}}
\def\IGa{\relax\hbox{${\rm I}\kern-.18em\Gamma$}}
\def\IH{\relax{\rm I\kern-.18em H}}
\def\IK{\relax{\rm I\kern-.18em K}}
\def\IL{\relax{\rm I\kern-.18em L}}
\def\IP{\relax{\rm I\kern-.18em P}}
\def\IR{\relax{\rm I\kern-.18em R}}
\def\Z{\relax\ifmmode\mathchoice
{\hbox{\cmss Z\kern-.4em Z}}{\hbox{\cmss Z\kern-.4em Z}}
{\lower.9pt\hbox{\cmsss Z\kern-.4em Z}} {\lower1.2pt\hbox{\cmsss
Z\kern-.4em Z}}\else{\cmss Z\kern-.4em Z}\fi}

\def\II{\relax{\rm I\kern-.18em I}}

\def\p{\partial}

%\SpradlinBN
\lref\SpradlinBN{ M.~Spradlin and A.~Strominger, ``Vacuum states
for AdS(2) black holes,'' JHEP {\bf 9911}, 021 (1999)
[arXiv:hep-th/9904143].
%%CITATION = HEP-TH 9904143;%%
}
%\MaldacenaUZ
\lref\MaldacenaUZ{ J.~M.~Maldacena, J.~Michelson and
A.~Strominger, ``Anti-de Sitter fragmentation,'' JHEP {\bf 9902},
011 (1999) [arXiv:hep-th/9812073].
%%CITATION = HEP-TH 9812073;%%
}

%\BrittoPacumioAX
\lref\screview{ R.~Britto-Pacumio, J.~Michelson, A.~Strominger and
A.~Volovich, ``Lectures on superconformal quantum mechanics and
multi-black hole  moduli spaces,'' arXiv:hep-th/9911066.
%%CITATION = HEP-TH 9911066;%%
}

\lref\gtt{S. Gukov, T. Takayanagi and N. Toumbas, to appear.}
%\gukov
\lref\gukov{ N.~Berkovits, S.~Gukov and B.~C.~Vallilo,
``Superstrings in 2D backgrounds with R-R flux and new extremal
black  holes,'' Nucl.\ Phys.\ B {\bf 614}, 195 (2001)
[arXiv:hep-th/0107140].
%%CITATION = HEP-TH 0107140;%%
}
\lref\sst{A. Simon, A. Strominger and D. Thompson in progress.}
%\MooreZV
\lref\MooreZV{ G.~W.~Moore, M.~R.~Plesser and S.~Ramgoolam,
``Exact S matrix for 2-D string theory,'' Nucl.\ Phys.\ B {\bf
377}, 143 (1992) [arXiv:hep-th/9111035].
%%CITATION = HEP-TH 9111035;%%
}

%\TakayanagiSM
\lref\TakayanagiSM{ T.~Takayanagi and N.~Toumbas, ``A matrix model
dual of type 0B string theory in two dimensions,'' JHEP {\bf
0307}, 064 (2003) [arXiv:hep-th/0307083].
%%CITATION = HEP-TH 0307083;%%
}

%\BoulatovXZ
\lref\bk{ D.~Boulatov and V.~Kazakov, ``One-dimensional string
theory with vortices as the upside down matrix oscillator,'' Int.\
J.\ Mod.\ Phys.\ A {\bf 8}, 809 (1993) [arXiv:hep-th/0012228].
%%CITATION = HEP-TH 0012228;%%
}

%\GrossAY
\lref\GrossAY{ D.~J.~Gross and N.~Miljkovic, ``A Nonperturbative
Solution of $D = 1$ String Theory,'' Phys.\ Lett.\ B {\bf 238},
217 (1990);
%%CITATION = PHLTA,B238,217;%%
}
%
%\GibbonsFA
\lref\gt{ G.~W.~Gibbons and P.~K.~Townsend, ``Black holes and
Calogero models,'' Phys.\ Lett.\ B {\bf 454}, 187 (1999)
[arXiv:hep-th/9812034].
%%CITATION = HEP-TH 9812034;%%
}

%\BrezinSS
\lref\BrezinSS{ E.~Brezin, V.~A.~Kazakov and A.~B.~Zamolodchikov,
``Scaling Violation in a Field Theory of Closed Strings in One
Physical Dimension,'' Nucl.\ Phys.\ B {\bf 338}, 673 (1990);
%%CITATION = NUPHA,B338,673;%%
}
%
%\GinspargAS
\lref\GinspargAS{ P.~Ginsparg and J.~Zinn-Justin, ``2-D Gravity +
1-D Matter,'' Phys.\ Lett.\ B {\bf 240}, 333 (1990).
%%CITATION = PHLTA,B240,333;%%
}
%\DasKA
\lref\DasKA{ S.~R.~Das and A.~Jevicki, ``String Field Theory And
Physical Interpretation Of D = 1 Strings,'' Mod.\ Phys.\ Lett.\ A
{\bf 5}, 1639 (1990).
%%CITATION = MPLAE,A5,1639;%%
}
%\DouglasUP
\lref\six{ M.~R.~Douglas, I.~R.~Klebanov, D.~Kutasov,
J.~Maldacena, E.~Martinec and N.~Seiberg, ``A new hat for the c =
1 matrix model,'' arXiv:hep-th/0307195.
%%CITATION = HEP-TH 0307195;%%
}

%\KazakovPM
\lref\kkk{ V.~Kazakov, I.~K.~Kostov and D.~Kutasov, `A matrix
model for the two-dimensional black hole,'' Nucl.\ Phys.\ B {\bf
622}, 141 (2002) [arXiv:hep-th/0101011].
%%CITATION = HEP-TH 0101011;%%
}
%\GrossUB
\lref\GrossUB{ D.~J.~Gross and I.~R.~Klebanov, ``One-Dimensional
String Theory On A Circle,'' Nucl.\ Phys.\ B {\bf 344}, 475
(1990).
%%CITATION = NUPHA,B344,475;%%
}
%\MartinecKA
\lref\MartinecKA{ E.~J.~Martinec, ``The annular report on
non-critical string theory,'' arXiv:hep-th/0305148.
%%CITATION = HEP-TH 0305148;%%
}
%\PolchinskiMB
\lref\PolchinskiMB{ J.~Polchinski, ``What is string theory?,''
arXiv:hep-th/9411028.
%%CITATION = HEP-TH 9411028;%%
}

%\KlebanovKM
\lref\kms{ I.~R.~Klebanov, J.~Maldacena and N.~Seiberg, ``D-brane
decay in two-dimensional string theory,'' JHEP {\bf 0307}, 045
(2003) [arXiv:hep-th/0305159].
%%CITATION = HEP-TH 0305159;%%
}
%\GinspargIS
\lref\GinspargIS{ P.~Ginsparg and G.~W.~Moore, ``Lectures On 2-D
Gravity And 2-D String Theory,'' arXiv:hep-th/9304011.
%%CITATION = HEP-TH 9304011;%%
}
%\McGreevyKB
\lref\hv{ J.~McGreevy and H.~Verlinde, ``Strings from tachyons:
The c = 1 matrix reloaded,'' arXiv:hep-th/0304224.
%%CITATION = HEP-TH 0304224;%%
}

\def\p{\partial}

\lref\mss{D. Thompson,"AdS2 Solutions of 2D Type OA"
hep-th/0312156.}

%\BrezinRB
\lref\BrezinRB{ E.~Brezin and V.~A.~Kazakov, ``Exactly Solvable
Field Theories Of Closed Strings,'' Phys.\ Lett.\ B {\bf 236}, 144
(1990).
%%CITATION = PHLTA,B236,144;%%
}
%\GrossVS
\lref\GrossVS{ D.~J.~Gross and A.~A.~Migdal, ``Nonperturbative
Two-Dimensional Quantum Gravity,'' Phys.\ Rev.\ Lett.\  {\bf 64},
127 (1990).
%%CITATION = PRLTA,64,127;%%
}
%\DouglasVE
\lref\DouglasVE{ M.~R.~Douglas and S.~H.~Shenker, ``Strings In
Less Than One-Dimension,'' Nucl.\ Phys.\ B {\bf 335}, 635 (1990).
%%CITATION = NUPHA,B335,635;%%
}
%\PolchinskiUQ
\lref\PolchinskiUQ{ J.~Polchinski, ``Classical Limit Of
(1+1)-Dimensional String Theory,'' Nucl.\ Phys.\ B {\bf 362}, 125
(1991).
%%CITATION = NUPHA,B362,125;%%
} \lref\KlebanovQA{ I.~R.~Klebanov, ``String theory in
two-dimensions,'' arXiv:hep-th/9108019.
%%CITATION = HEP-TH 9108019;%%
}

%\NatsuumeSP
\lref\NatsuumeSP{ M.~Natsuume and J.~Polchinski, ``Gravitational
Scattering In The C = 1 Matrix Model,'' Nucl.\ Phys.\ B {\bf 424},
137 (1994) [arXiv:hep-th/9402156].
%%CITATION = HEP-TH 9402156;%%
}

%\PolchinskiJP
\lref\PolchinskiJP{ J.~Polchinski, ``On the nonperturbative
consistency of d = 2 string theory,'' Phys.\ Rev.\ Lett.\  {\bf
74}, 638 (1995) [arXiv:hep-th/9409168].
%%CITATION = HEP-TH 9409168;%%
}

%\AlexandrovCM
\lref\AlexandrovCM{ S.~Alexandrov and V.~Kazakov, ``Correlators in
2D string theory with vortex condensation,'' Nucl.\ Phys.\ B {\bf
610}, 77 (2001) [arXiv:hep-th/0104094].
%%CITATION = HEP-TH 0104094;%%
}

%\AlexandrovFH
\lref\AlexandrovFH{ S.~Y.~Alexandrov, V.~A.~Kazakov and
I.~K.~Kostov, ``Time-dependent backgrounds of 2D string theory,''
Nucl.\ Phys.\ B {\bf 640}, 119 (2002) [arXiv:hep-th/0205079].
%%CITATION = HEP-TH 0205079;%%
}

%\AlexandrovPZ
\lref\AlexandrovPZ{ S.~Y.~Alexandrov and V.~A.~Kazakov,
``Thermodynamics of 2D string theory,'' JHEP {\bf 0301}, 078
(2003) [arXiv:hep-th/0210251].
%%CITATION = HEP-TH 0210251;%%
}

%\AlexandrovQK
\lref\AlexandrovQK{ S.~Y.~Alexandrov, V.~A.~Kazakov and
I.~K.~Kostov, ``2D string theory as normal matrix model,'' Nucl.\
Phys.\ B {\bf 667}, 90 (2003) [arXiv:hep-th/0302106].
%%CITATION = HEP-TH 0302106;%%
\lref\jlkas{} }
%\BerkovitsTG
\lref\BerkovitsTG{ N.~Berkovits, S.~Gukov and B.~C.~Vallilo,
%``Superstrings in 2D backgrounds with R-R flux and new extremal black  holes,''
Nucl.\ Phys.\ B {\bf 614}, 195 (2001) [arXiv:hep-th/0107140].
%%CITATION = HEP-TH 0107140;%%
}
%\BanksXS
\lref\banks{ T.~Banks and M.~O'Loughlin, ``Nonsingular Lagrangians
for two-dimensional black holes,'' Phys.\ Rev.\ D {\bf 48}, 698
(1993) [arXiv:hep-th/9212136].
%%CITATION = HEP-TH 9212136;%%
}

%\draft
%-------------------
% title page
%-------------------
%
\Title{\vbox{\baselineskip12pt \hbox{hep-th/0312194}}} {\vbox{
\centerline {A Matrix Model for AdS$_2$}}} \centerline{Andrew
Strominger} \vskip.1in {\it Jefferson Physical Laboratory, Harvard
University, Cambridge, MA 02138}

\vskip.1in \centerline{\bf Abstract} {A matrix quantum mechanics
with potential $V={q^2 \over r^2}$ and an SL(2,R) conformal
symmetry is conjectured to be dual to two-dimensional type 0A
string theory on AdS$_2$ with $q$ units of RR flux.}

 \Date{}

\listtoc\writetoc
\newsec{Introduction}

The reinterpretation \refs{\hv \kms \TakayanagiSM- \six} of $c=1$
matrix models as holographic duals of type 0 string theories has
revitalized two-dimensional string theories as useful toy models
for understanding their higher-dimensional cousins. In this paper
we will use this to motivate a conjecture for an AdS$_2$/CFT$_1$
correspondence, relating an AdS$_2$ solution of the type 0A
effective action to a conformally invariant matrix quantum
mechanics.

Type 0A string theory in two dimensions has a single scalar
"tachyon" as well as non-dynamical RR two-form fluxes \six. The
low energy effective action with $q$ units of electric flux has
extremal black hole-like solutions with a near-horizon AdS$_2$
region \refs{\banks,\gukov}. The string coupling in the AdS$_2$
region can be made arbitrarily weak by making $q$ large. We argue
in section 2 that the extremal solutions comprise a one parameter
family of solutions labelled by the thickness $\mu$ of the
"tachyon hair" . However in the AdS$_2$ region the curvature is of
order one in string units, so results from the effective action
are at best suggestive. In this paper we shall simply assume  the
existence of a tree-level 0A AdS$_2$ solution.\foot{Since the
string coupling can be made arbitrarily weak, the existence of
such a region is a question for the string worldsheet, potentially
answerable by an analysis of RR CFTs.}

A holographic dual for the 2D 0A string theory has been proposed
\six\ as the quantum mechanics of free fermions moving in the
potential \eqn\dcl{V=-{r^2\over 4\alpha '}+{q^2 \over 2 r^2}.}
This has a one parameter family of static solutions labelled by
the energy $\mu$ of the fermi surface. We propose to identify this
parameter with the thickness of the tachyon hair of the extremal
black holes in the spacetime picture.

At very small $r$, the potential \dcl\ is approximated by
\eqn\llo{V={q^2 \over 2 r^2}.} We conjecture in section 3 that the
free fermion theory with potential \llo\ is dual to the purported
AdS$_2$ solution of 0A. The main piece of evidence for this is
that \llo\ defines an SL(2,R) conformally invariant quantum
mechanics which matches the SL(2,R) isometries of AdS$_2$.

This conjecture relates a well-defined matrix model to a 0A
spacetime solution which (unfortunately) has not been constructed
in the full tree-level string theory. Therefore it is difficult at
this point to find evidence for the conjecture, beyond matching
the symmetries. We do explain in section 4 how to use the matrix
model to compute the mass of the tachyon in AdS$_2$ which might,
at some later date, be compared to a worldsheet computation.

If correct our proposal has conceptual implications for the nature
of time in the matrix model (see also \gt). We argue in section 5
that the usual matrix model hamiltonian $H$ generates evolution
with respect to Poincare time in AdS$_2$. Evolution in global time
is generated by $L_0=\half(H+K)$, where $K$ is the generator of
special conformal transformations. The spectrum of $H$ is
continuous while that of $L_0$ is discrete. Hence different
operators in the free fermion theory correspond to hamiltonians
which evolve along different time slicings.

Past treatments of matrix models have largely rigidly invoked a
single hamiltonian. This is antithetical to the spirit of gravity
in which there are many time slicings and many hamiltonians. We
expect that quite generally, as in the example of this paper,
different operators in the various matrix models will have
illuminating interpretations as different hamiltonians.

\newsec{Spacetime solutions of $c=1$ 0A string theory}

\subsec{An extremal black hole} The low-energy effective action of
0A string theory in two dimensions, with RR electric field
proportional to $q$  and $\alpha^\prime=1$, is
\refs{\six,\mss}\foot{The low energy theory contains two $U(1)$
gauge fields $F^+$ and $F^-$ which give rise to opposite-signed
tachyon tadpoles. The action given here is for the sector of the
theory in which both the electric flux in both sectors is
proportional to $q$, and the tachyon tadpole is cancelled. $q$ is
the net number of D0brane sources up to factors of 2 and
$\pi$.}\eqn\seff{S_{eff}=\int d^2x
\sqrt{-g}\bigl(e^{-2\Phi}\bigl(8+R +4(\nabla \Phi)^2-\half (\nabla
T)^2 +T^2)-{q^2\over 2}-{q^2T^2}+.....\bigr).} Static, extremal
black hole-like solutions with $T=0$ of this (and more general
actions) have been found in \refs{\banks ,\gukov, \gtt}
\eqn\wxef{\eqalign{ds^2&=(1+{q^2\over 8}(\Phi-\Phi_0-\half)
e^{2\Phi} )(-dt^2+d\sigma^2),\cr
          \sigma(\Phi) &={1 \over \sqrt{2}}\int^{\Phi}{d\Phi' \over 1+{q^2\over
          8}(\Phi'-\Phi_0-\half)
          e^{2\Phi'}},\cr
          \Phi_0 &\equiv -\ln{q \over 4 }.}}
In the asymptotic region $\sigma \to -\infty$ (or $\sigma << -\ln
q$), the coupling becomes weak, the effect of the flux $q$ is
diminished, and the solution approaches the usual linear dilaton
vacuum \eqn\wef{\eqalign{ds^2&\to -dt^2+d\sigma^2,\cr
          \Phi &= \sqrt{2}\sigma,~~~~~~~~~~~~~~\sigma  \to -\infty.}}
In the "near-horizon" region $\sigma \to +\infty$ ($\sigma>> \ln
q$), $\Phi$ reaches the critical value $\Phi=\Phi_0$ and the
conformal factor of the metric in \wxef\ vanishes quadratically
\eqn\vzh{\eqalign{ds^2&\to - 2(\Phi-\Phi_0)^2(dt^2-d\sigma^2),\cr
\Phi&\to \Phi_0-{1 \over 2\sqrt{2} \sigma}~~~~\sigma \to +\infty
.}} The Killing vector $\p_t$ becomes null, so we identify
$\sigma=\infty$ as a horizon. The near-horizon geometry is
\eqn\rfc{\eqalign{ds^2&= {1 \over 4\sigma^2}(-dt^2+d\sigma^2),\cr
          \Phi &=\Phi_0~~~~~~~ ~~~~~~~~~~~~~~~~\sigma  \to +\infty.}}
This is easily recognized as the $SL(2,R)$ invariant metric on the
Poincare patch of AdS$_2$. The string coupling $e^{\Phi_0}={4\over
q}$ will be small if $q$ is large.\foot{We note that this
$q$-dependence agrees with a matrix model determination of the
effective string coupling in \gtt.} On the other hand the scalar
curvature is \eqn\scft{R=-8,~~~~\ell_{AdS}=\half} in string units.
This means that, although string loops are suppressed by large
$q$, $\alpha^\prime$ corrections remain large and \seff\ cannot be
trusted. For example $R^2$ or $R^2T^2$ terms are important.
Nevertheless we shall proceed by assuming that the suggestion of a
near-horizon AdS$_2$ region is correct, and that \rfc\ corresponds
to an AdS$_2$ solution of 0A string theory at weak string
coupling. In principle this might be verified from a study of
string worldsheet CFTs with RR backgrounds, although with our
current level of understanding of RR backgrounds this may not be
an easy task.

The full solution \wxef\ hence describes the two dimensional
reduction of an extremal black hole interpolating from a flat
linear dilaton region at infinity to an AdS$_2$ near-horizon
region. The present paper largely concerns the AdS$_2$ region on
its own. However there are infrared subtleties in the physics of
AdS$_2$ \refs{\MaldacenaUZ, \SpradlinBN} which suggest that the
decoupling of the near horizon and asymptotic regions may not be
as simple as it is for the higher-dimensional analogs. Further
progress in this area will likely require a better understanding
of these issues.

 There are also non-extremal analogs of the solutions \wxef\ \gukov\
which have a finite Hawking temperature and are quantum
mechanically unstable. These solutions differ by an integration
constant (the mass parameter) to the conformal factor in \wxef\
and have a single rather than a double zero at the horizon.
However the focus of this paper will be exactly static stable
solutions and these non-extremal solutions will accordingly not be
considered.

\subsec{Tachyon Hair}

When $q=0$, \wxef\ reduces to the familiar linear dilaton vacuum.
This solution has a one-parameter (usually denoted by $\mu$)
family of static deformations. In the asymptotic region the
solution has a nonzero tachyon field decaying as a certain linear
combination of $\mu \sigma e^{\sigma}$ and $\mu e^{\sigma}$. In
this subsection we discuss the generalization of this family of
solutions to nonzero $q$.\foot{Black hole no-hair theorems do not
apply here because of the weak boundary conditions at infinity.}

The linearized tachyon equation of motion following from $S_{eff}$
is \eqn\ddfc{\nabla^2(e^{-\Phi}T) +(2-(\nabla \Phi)^2
+\nabla^2\Phi-{q^2 \over
 \pi}e^{2\Phi)}) (e^{-\Phi}T) =0.}
 In the asymptotic region, for
time-independent $T$ this becomes
\eqn\dxvm{-\p_\sigma^2(e^{-\sqrt{2}\sigma}T) = 0,} which leads to
the general asymptotic solution\foot{We find it convenient in the
current context to adopt conventions in which $\mu$ is a positive,
rather than negative, multiple of the energy.} \eqn\tam{T\sim -
(\mu\sigma + a ) e^{\sqrt{2}\sigma} ,~~~~\sigma \to -\infty .} In
the near-horizon AdS$_2$ region the equation becomes
\eqn\dpo{e^{-\Phi_0}(\nabla^2 T-m_T^2 T)=0,} \eqn\dzv{m_T^2=30,}
whose general static solution is \eqn\nhr{T\sim b_- \sigma^{h_-}+
b_+\sigma^{h_+}~~~~~\sigma \to +\infty,} where \mss
\eqn\rgy{h_\pm=\half(1\pm\sqrt{1+4m_T^2\ell_{AdS}^2}) .} We expect
the actual values of $h_\pm$ to be affected by $\alpha^\prime$
corrections, but we will assume that $h_-$ is zero or negative. We
will see below that the matrix model correspondence predicts
definite values for $h_\pm$. Knowledge of the exact solution of
\ddfc, which could be solved numerically, would relate the two
parameters $(\mu, a)$ to $(b_-,b_+)$, but the numerical
coefficient will not be needed for our purposes. Regularity on the
horizon at $\sigma=\infty$ requires that $b_+=0$, which in turn
fixes $a$ as a linear function of $\mu$.

To summarize, the action \wxef\ has a one-parameter family of
smooth, static, quantum-mechanically stable solutions which may be
described as extremal black holes with tachyon hair. The parameter
can be taken to be the coefficient $\mu$ of the exponential tail
of the tachyon at spatial infinity. These solutions generalize the
usual tachyon wall solutions to nonzero $q$.

\newsec{Static solutions of the 0A matrix model}

Type $0A$ theory with $D0$-brane flux proportional to $q$ in two
dimensions is described by the matrix quantum mechanics. This can
be expressed as a theory of free fermions $\Psi(x)$ moving on the
plane in the radially symmetric potential\six\foot{There is a
potentially important subtlety here which we have not resolved. In
 \six\ this potential is argued to be relevant to the case of only
one type of zero brane charge, for which the spacetime effective
action has a tachyon tadpole \mss. It is not clear in this case
how the tadpole is cancelled: in the (not fully reliable)
effective action \seff\ we cancel this tadpole by turning on the
second charge. Therefore it is possible that an enlarged (due to
the second type of D0branes) matrix model could be relevant to the
AdS$_2$ spacetime solutions.

To be precise $q^2$ in the expressions of this section should be
replaced by $q^2-{1 \over 4}$ but this difference, as well as
convention dependent numerical factors relating this $q$ to the
one of the previous section, are unimportant here and will be
suppressed.} \eqn\ftd{V=-{r^2 \over 4}+{q^2 \over 2 r^2}.} The
fermions are constrained to have zero angular momentum. This
theory has a one parameter family of solutions which can be simply
described as filling all states in the fermi sea up to energy
$\mu$. We conjecture that this parameter $\mu$ is (proportional to
) the coefficient $\mu$ parameterizing the tachyon hair of the
previous section.\foot{The special case $\mu=0$ of this conjecture
is due to the authors of \gtt.} Let us try to see how this fits
together for $q\neq 0$ in the 0A theory.

For large $q$ the quantum mechanics has two dynamical regions,
divided around $r^2\sim q$,  where the potential is dominated by
the $-r^2$ or $q^2 \over r^2$ terms. Deep in the first region we
can ignore the $q^2 \over r^2$ term. Hence this can be identified
as the linear dilaton region. In the second region $r^2<<q$ the
free fermion hamiltonian is approximately \eqn\oork{H =\half\int
d^2x\Psi^\dagger \bigl(p^2+{q^2 \over r^2}\bigr)\Psi,} where
$p^2=- {1 \over r}\p_r r\p_r$ on radially symmetric wavefunctions.
It is natural to identify this as the AdS$_2$ region.\foot{The
precise relation between $r$ and the spacetime coordinate $\sigma$
could be complicated due to leg pole factors. Our identification
of the matrix model and spacetime quantities will not require this
relation, but will proceed (see below) via the SL(2,R)
symmetries.} We conjecture that \oork\ on its own describes string
theory in AdS$_2$.

An important check on this conjecture is provided by the
symmetries. \oork\ turns out to be conformally invariant. The
$SL(2,R)$ symmetry is generated by $H$ together with
\eqn\dsft{K=\half\int d^2x\Psi^\dagger r^2\Psi~~~~~~~~~~~ D=\half
\int d^2x\Psi^\dagger (r p_r+p_rr) \Psi,} where here $p_r=-i\p_r$.
The commutators are \eqn\hhi{[D,H]=2iH,~~~[D,K]=-2iK,~~~[H,K]=-iD.
}This $SL(2,R)$ matches the $SL(2,R)$ isometries of the gravity
AdS$_2$ solution, which are generated by,  in the metric \rfc,
\eqn\sltr{H=i\p_t,~~~~~D=-2i(\sigma\p_\sigma+t\p_t),
~~~~~K=i(2t\sigma\p_\sigma+t^2\p_t+\sigma^2\p_t).} This is the
main piece of evidence for the conjecture. If correct the duality
relating the quantum mechanics described by \oork\ to 0A string
theory with $q$ units of flux in the geometry \rfc\ is an example
of an $AdS_2/CFT_1$ correspondence.\foot{We wish to draw attention
to a possibly relevant set of conformal quantum theories known as
the Calogero-Moser models and considered in a related context in
\gt. These are obtained by replacing the eigenvalue potential of
the AdS$_2$ matrix model \eqn\wer{V=\sum_{i=1}^N {q^2 \over
\lambda_i^2}} with \eqn\wser{V=\sum_{i<j}^N {q^2 \over
(\lambda_i-\lambda_j)^2}.} Similar quantum-mechanical models arise
in the study of black hole moduli spaces \screview. \wser\ gives
an integrable, conformally invariant quantum theory. Hence, as
pointed out in \gt, it is a natural candidate for the $CFT_1$ dual
to an AdS$_2$ string theory. In this context it is intriguing to
note that exactly this theory arises in the non-singlet sector of
the $c=1$ matrix model \bk. This may be the proper setting for the
spacetime dual to the Calogero-Moser sought in \gt.}

Next we consider the static solutions with fermi surfaces at
energies $\mu$. These correspond to quantum states $|\mu\rangle$
obeying \eqn\fll{H|\mu>=\mu|\mu\rangle.}For large negative $\mu$,
a small wave coming in from spatial infinity on the fermi sea is
reflected long before it reaches the AdS$_2$ region. In the
gravity picture this is because large negative (with our
conventions) $\mu$ corresponds to a large positive tachyon wall
shielding the AdS$_2$ region. On the other hand for large positive
$\mu$ the wave will travel deep into the AdS$_2$ region before
being reflected.  This is because positive $\mu$ corresponds to
negative tachyon field which goes to zero at spatial infinity and
the AdS$_2$ horizon, creating a tachyon ditch which sucks the
waves across the potential barriers separating the linear dilaton
and AdS$_2$ regions. These potential barriers are the same as
those that lead to greybody factors in black hole evaporation. For
large $\mu$ and deep in the AdS$_2$ region the fermi surface obeys
\eqn\fsc{p^2+{q^2 \over r^2}=\mu.} The nonzero eigenvalue for $H$
corresponds to a spontaneous breaking of the SL(2,R) symmetry.
\newsec{The tachyon mass in AdS$_2$}

Due to the lamentable current status of our technology for
understanding RR worldsheet CFTs , we do not know if there is a
AdS$_2$ worldsheet CFT for the $0A$ theory, let alone how to
compute the spectrum. This makes it difficult to find any checks
of our conjecture. However, assuming the conjecture, we can make a
prediction for the tachyon mass (in units of the AdS$_2$
curvature)  in this CFT  as follows.

The dilation operator $D$ rescales the hamiltonian according to
\eqn\drf{e^{-i\alpha D}H e^{i\alpha D}=e^{2\alpha}H .}Hence
\eqn\wyu{e^{i\alpha D}|\mu \rangle =| e^{2\alpha}\mu\rangle.} On
the other hand in the AdS$_2$ solution the dilation operator is
the Killing vector \eqn\ffa{D=-2i\sigma\p_\sigma -2it\p_t.}
 This acts
on the near horizon tachyon field \nhr\ (with $b_+=0$) as
\eqn\retl{e^{-i\alpha D}T=e^{-2\alpha h_-}T.}Recalling that $T$ is
itself proportional to $\mu$ this is equivalent to \eqn\mio{\mu
\to e^{-2\alpha h_-}\mu.} Comparing \mio\ to \wyu\ leads to
\eqn\dool{h_-=-1,} or equivalently a tachyon mass
\eqn\reft{m_T^2={2\over \ell_{AdS}^2},} where $\ell_{AdS}$ is the
AdS$_2$ radius. This may be viewed as a prediction for the
spectrum of the worldsheet RR CFT .

\newsec{Time in the matrix model}
The conjectured AdS$_2$/CFT$_1$ correspondence has implications
for the nature of time in the matrix model. The following
discussion parallels that given in a related context by Gibbons
and Townsend \gt.

The hamiltonian $H$ in \oork\  generates time translations $i\p_t$
in the Poincare patch of AdS$_2$ with metric
\eqn\met{ds^2={-dt^2+d\sigma^2 \over 4 \sigma^2}.} This patch
covers only a wedge-shaped region of AdS$_2$. Global coordinates
can be defined by \eqn\fty{\tau \pm w =2 {\rm arctan }(t\pm
\sigma),} in terms of which the metric is
\eqn\dki{ds^2={-d\tau^2+dw^2 \over 4\sin ^2w}.} Translation in
global time $\tau$ are generated  by \eqn\fto{L_0\equiv i\p_\tau
=\half(H+K).} Identifying the spacetime Killing vectors \sltr\
with the matrix model operators \oork,\dsft, $L_0$ corresponds to
a matrix model hamiltonian \eqn\ork{L_0 =\half\int dx\Psi^\dagger
\bigl(p^2+{q^2 \over r^2}+r^2\bigr)\Psi.} Unlike $H$, $L_0$ has a
potential which grows in both directions and a corresponding
discrete spectrum. This is in agreement with the expected discrete
spectrum of string theory in AdS$_2$.
 The $SL(2,R)$ invariant
ground state is the state with no fermions in the coordinates
\dki. $L_0$ is naturally written as part of a SL(2,R) multiplet
\eqn\dcxx{L_0=\half(H+K),~~~~~L_{\pm1}=\half(H-K\mp iD)} obeying
\eqn\vvv{[L_0,L_{\pm 1}]=\mp L_{\pm
1},~~~~~~~~~~~[L_1,L_{-1}]=2L_0.} It is natural to characterize
the AdS$_2$ matrix model in terms of $L_0$ rather then $H$
eigenstates. In the gravity picture, this simply corresponds to a
different choice of time .

There is potentially a general lesson for matrix models here. In
the spacetime picture, they are theories of gravity. Theories of
gravity have many equivalent descriptions corresponding to
different time slicings and different hamiltonians. Often it is
necessary to consider a variety of time slicings to fully
understand the physics. The usual description of matrix models
involves a single fixed hamiltonian $H$. Our current perspective
suggests that some operators other than the usual $H$ in the
matrix model should correspond to different hamiltonians and
different time slicings.

\centerline{\bf Acknowledgements}
 This work was supported in part by DOE grant DE-FG02-91ER40654. I am grateful to
 J. Maldacena, S. Minwalla, J. Polchinski, N.
Seiberg, S. Shenker, D. Thompson, X. Yin and especially J.
Karczmarek and T. Takayanagi for useful conversations.

\listrefs

\end